\documentstyle[aps,prl,epsfig,manuscript,draft]{revtex}

\begin{document}

{
\title{New Bounds from Searching for Muonium to Antimuonium Conversion}
\author{
L.~Willmann$^1$,
P.V.~Schmidt$^1$,
H.P.~Wirtz$^2$,
R.~Abela$^3$,
V.~Baranov,$^4$,
J.~Bagaturia$^5$,
W.~Bertl$^3$,
R.~Engfer$^2$,
A.~Gro\ss{}mann$^1$,
V.W.~Hughes$^6$,
K.~Jungmann$^1$,
V.~Karpu\-chin$^4$,
I.~Kisel$^4$,
A.~Korenchenko$^4$,
S.~Korenchenko$^4$,
N.~Kravchuk$^4$,
N.~Kuchinsky$^4$,
A.~Leuschner$^2$
V.~Meyer$^1$,
J.~Merkel$^1$,
A.~Moiseenko$^4$,
D.~Mzavia$^5$,
G.~zu~Putlitz$^1$,
W.~Reichart$^2$,
I.~Reinhard$^1$,
D.~Renker$^3$,
T.~Sakhelash\-vil\-li$^5$,
K.~Tr\"ager$^1$ and
H.K.~Walter$^3$
}
\address{
$^1$ Physikalisches Institut, Universit\"at
Heidelberg, D-69120 Heidelberg, Germany
\newline
$^2$ Physik Institut, Universit\"at Z\"urich, CH-8057 Z\"urich, Switzerland
\newline
$^3$ Paul Scherrer Institut, CH-5232 Villigen PSI, Switzerland
\newline
$^4$ Joint Institute of Nuclear Research, RU-141980 Dubna, Russia
\newline
$^5$ Tbilisi State University, GUS-380086 Tbilisi, Georgia
\newline
$^6$ Physics Department, Yale University, New Haven Ct., 06520, USA }
\maketitle
\begin{abstract}
{
A new upper limit for the
probability of spontaneous muonium to antimuonium conversion
was established at ${\rm P_{M\overline{M}}}
\leq 8.2 \cdot 10^{-11}$ (90\%C.L.) in 0.1~T magnetic field,
which implies consequences for speculative extensions to the standard model.
Coupling parameters in R-parity violating supersymmetry
and the mass of a flavour diagonal bileptonic gauge boson can be
significantly restricted.
A Z$_8$ model with radiative mass generation through heavy lepton seed and
the minimal version of 331-GUT models are ruled out.
}
\end{abstract}
\vspace{5mm}

}
%

%
\newpage
At present all confirmed experimental experience is in agreement
with conserved lepton numbers. Several solely empirical laws
appear to hold simultaneously including
multiplicative and additive schemes \cite{cabb61}.
No associated symmetry has yet been identified thus
leaving lepton numbers in a unique status in physics,
since flavour mixing in the quark sector
is well established and described by the Cabbibo-Kobayashi-Maskawa
matrix.
The standard model in particle physics assumes
additive lepton family number conservation, and any observed violation
would be a clear indication of new physics.
In  many speculative theories, which
extend the standard model in order to
explain some of its
features such as parity violation in the weak interactions
or CP violation, lepton flavours are not conserved.
These theories have motivated a variety of dedicated sensitive
searches for rare decay modes of muons and kaons \cite{coop97}
and for neutrino oscillations.
%

%
%

%

\begin{figure}[h]
     \centering{
      \epsfig{figure=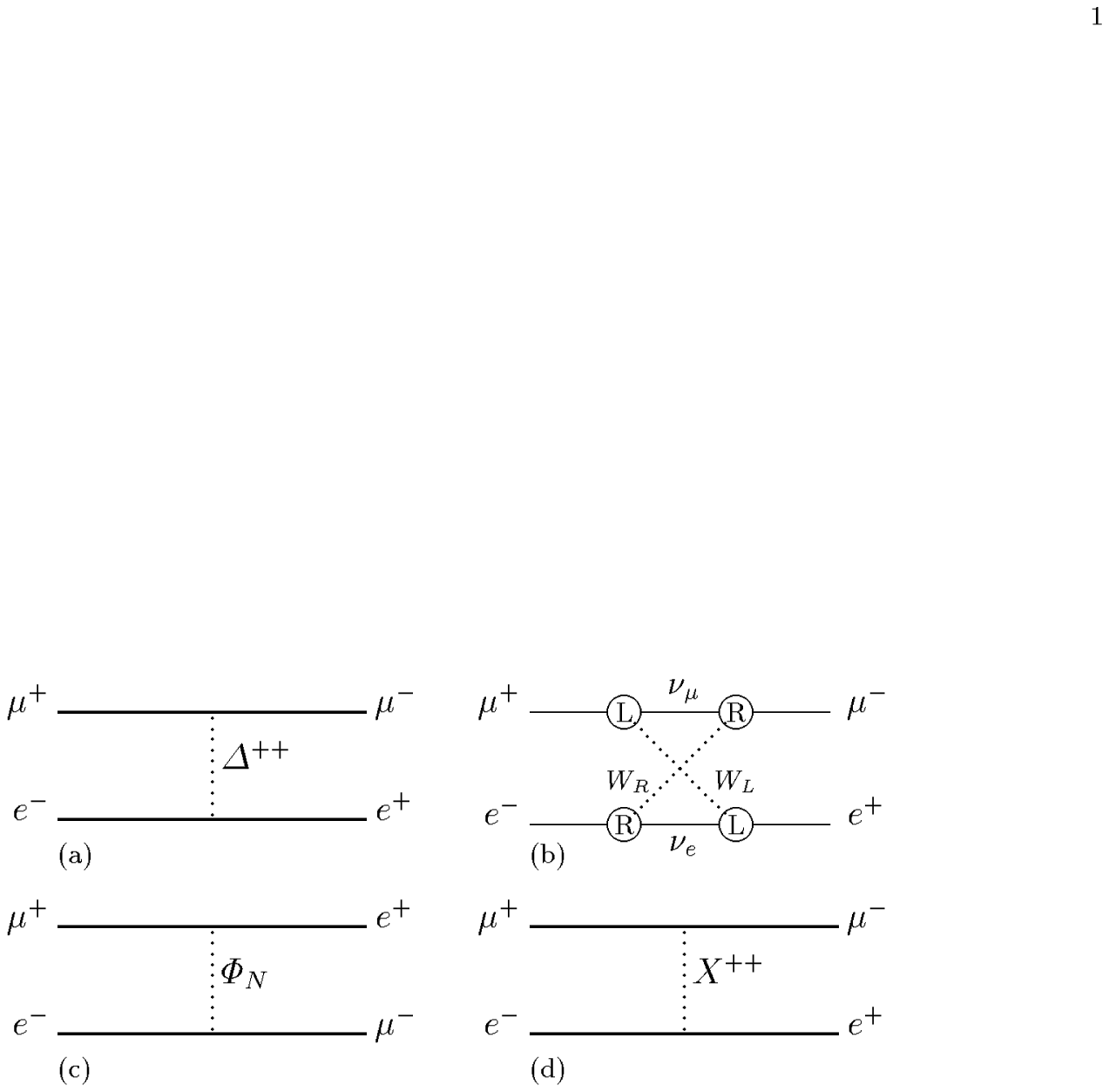,width=3.2in}
       \caption[thb]
       {\protect{\label{theo_mmb}}
       Muonium-antimuonium conversion in
       theories beyond the standard model. The interaction
       could be mediated, e.g. by
       (a) doubly charged Higgs boson $\Delta^{++}$
       \protect{\cite{halp82,herc92}},
       (b) heavy Majorana neutrinos \protect{\cite{halp82}},
       (c) a neutral scalar $\Phi_N$ \protect{\cite{wong94}}, e.g.
       a supersymmetric $\tau$-sneutrino $\tilde{\nu}_{\tau}$
       \protect{\cite{moha92,halp93}}, or
       (d) a bileptonic flavour diagonal gauge boson
       $X^{++}$ \protect{\cite{fuji94,fram97}}.
       }
              }
\end{figure}

 Of particular interest is the muonium atom (M=$\mu^+ e^-$)
which consists of two leptons from different generations.
As the electromagnetic part of the binding is well
described by electroweak standard theory it renders the possibility
to determine accurate values of fundamental constants as well
as searching for additional, yet unrevealed electron-muon interactions.
A spontaneous conversion of muonium into antimuonium
(${\rm \overline{M}} = \mu^- e^+)$ would violate additive lepton family
number conservation by two units, however, it is allowed by a multiplicative law.
This process could play a decisive role in
many speculative models (Fig. \ref{theo_mmb})
\cite{herc92,halp82,wong94,moha92,halp93,fuji94,fram97}.

The measurements reported here were performed with
the Muonium - Antimuonium - Con\-ver\-sion - Spec\-trome\-ter (MACS)
whose design is based on the observation of
M atoms in vacuo. In matter the possible conversion is strongly
suppressed mainly
due to the loss of symmetry between M and
${\rm \overline{M}}$ due to the possibility of $\mu^-$ transfer in collisions involving
${\rm \overline{M}}$ \cite{morg70,fein61}.
The required signature of a conversion process is the coincident
identification of both the electron and positron released in the decay of
the antiatom \cite{matt91,abel96}. An energetic electron ($e^-$) arises from the decay
$\mu^- \rightarrow e^- + \nu_{\mu} + \overline{\nu}_e$ with a characteristic Michel
energy distribution extending to 53 MeV \cite{Mich50}, and a positron ($e^+$)
appears with an average
kinetic energy of 13.5 eV corresponding to its momentum distribution
in the atomic 1s state of ${\rm \overline{M}}$ \cite{chat92} .

\begin{figure}[htb]
 \centering
     \epsfig{file=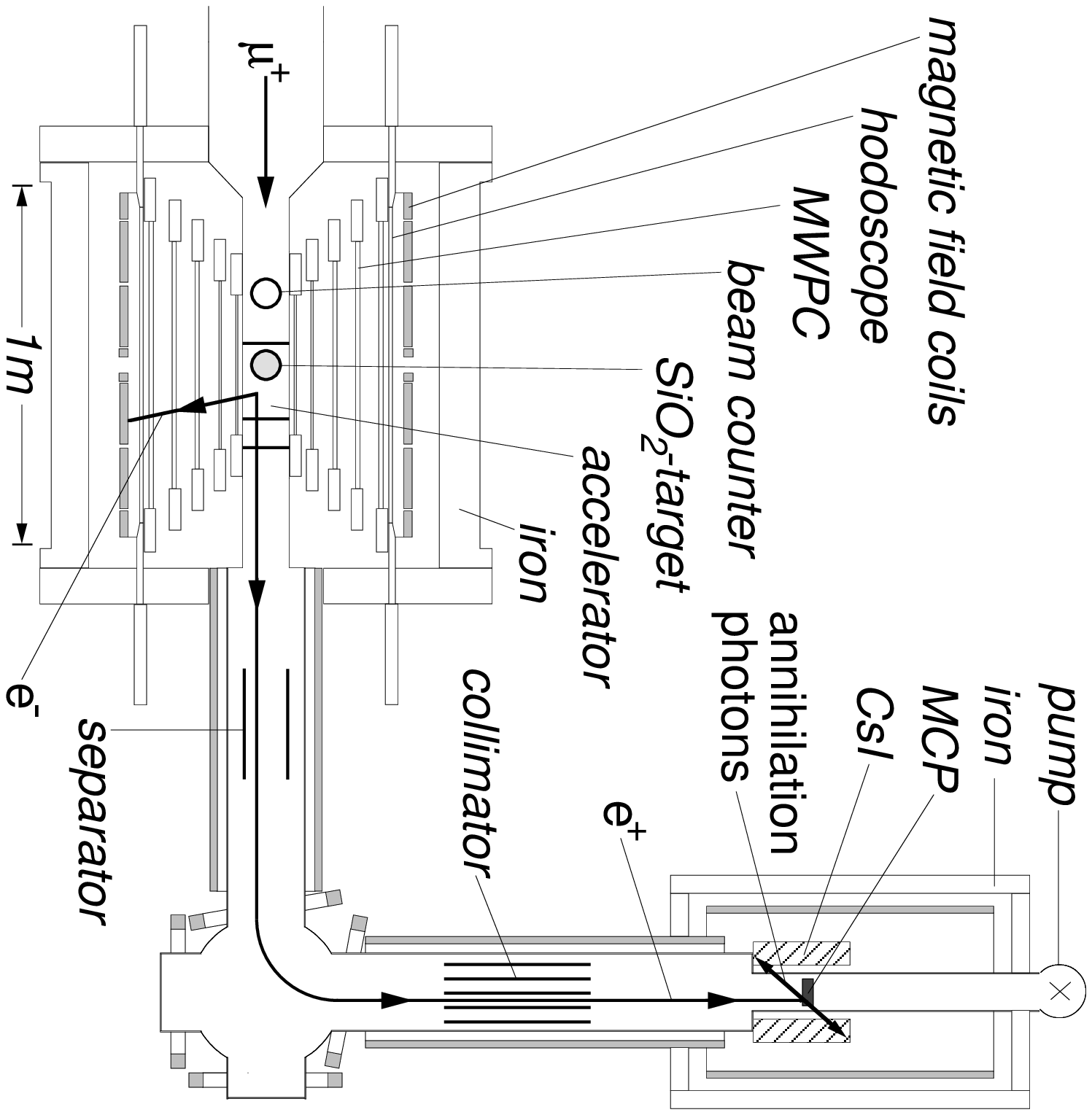,width=3.0in,angle=90}
      \caption[]{
      \protect{\label{mmbarsetup}}
      The MACS apparatus
      at PSI searching for
      muonium-antimuonium conversion.
      The signature requires observation of the energetic
      $e^-$ from $\mu ^-$ decay
      of ${\rm \overline{M}}$ in a magnetic spectrometer,
      in coincidence with the atomic shell $e^+$.
      The latter is accelerated and magnetically guided
      onto a microchannel plate (MCP). In addition at least one
      annihilation photon is required in a CsI calorimeter.
                 }
\end{figure}

The setup
has a large acceptance for these charged final state particles
(Fig. \ref{mmbarsetup}).
Its symmetry for detecting M and ${\rm \overline{M}}$ decays
through reversing all electric and magnetic fields is exploited in
regular measurements of the M atom production yield which is required for
normalization and, in addition, for monitoring
detector performance.
As a particular advantage, systematic uncertainties arising from corrections for
efficiencies and acceptances of various detector components cancel out.

The experiment utilizes the
world's brightest continuous surface muon channel $\pi$E5 \cite{abe92} at the Paul
Scherrer Institut (PSI) in Villigen, Switzerland. It provides a central
momentum  p=26 MeV/c, a momentum bite $\Delta$p/p=5\%~ and
rates up to 8$\cdot$ 10$^6$$\mu^+$/s.
The beam passes through a 280~$\mu$m scintillation counter
and a 270~$\mu$m mylar degrader. Muonium atoms are formed by electron capture
with 60\% yield after stopping the $\mu^+$ in a SiO$_2$ powder target
of thickness 8 mg/cm$^2$ and supported in vacuo
by a 25~$\mu$m aluminum foil with 30$^\circ$ inclination with respect
to the muon beam axis.
Most of the atoms emerge from the powder grains into the intergranular voids
and then on average 3.3 \% of them leave the target
surface with thermal Maxwell-Boltzmann velocity distribution
at 300~K \cite{wood88}.

When searching for ${\rm \overline{M}}$ decays the energetic $e^-$
is detected in a magnetic spectrometer
operated at 0.1~T magnetic field and covering 0.73$*4\pi$ solid angle
around the M production target. It has five concentric multiwire
proportional chambers with
radii of 8.2 to 32.0 cm and  active
lengths of 38 to 80 cm.
They all are equipped with two planes of segmented helical cathode stripes
for measuring radial, angular and axial coordinates.
The efficiency for full particle track
reconstruction is 82(2) \% and 90(2)\% if axial information is not requested.
The momentum resolution at 50 MeV/c is 54(2) \% yielding  a probability of $10^{-5}$
for misidentifying the charge of the particle.
It is limited by the 2~mm wire spacing in the proportional chambers.
The chambers are surrounded by a 64-fold segmented hodoscope providing a time
resolution of better than 1~ns.
Subsequent to
the $\mu^-$ decay the atomic shell $e^+$
is accelerated to typically 7~keV in a two stage electrostatic
device.
It is guided in an axial 0.1~T magnetic field along
a 3~m long transport region onto a microchannel plate detector (MCP)
with resistive anode readout and 0.55(3)~mm intrinsic position resolution.
A 35~mg/cm$^2$ magnesium oxide coated carbon foil in front of a commercial MCP
triplet stack serves as an additional
secondary electron emitter and yields a fourfold  enhancement of the detection
efficiency to 64(2)\% and reduces background counts of
low energy ions trapped in the magnetic field \cite{schmidt96}.
 The transport system is momentum selective due to
 a 90$^\circ$ horizontal bend of radius 35~cm  and a collimator consisting of
 40~cm long, 1~mm thick and 9~mm separated copper sheats which act to
suppress particles with longitudinal momenta exceeding 750~keV/c
because their gyration radii exceed 4.5~mm in the magnetic guiding field.
 The field gradient in the bend region
causes a vertical drift for charged particles proportional to their momenta.
 It is compensated for 7~keV $e^+$ by a 270~V/cm transverse electrostatic
field region of length 40~cm preceding the bend which also deflects
low energy $\mu^+$ and ions.
%

Positrons are uniquely identified by annihilation radiation when striking the MCP
which is centered inside of a barrel shaped 12-fold segmented barrel shaped pure
CsI crystal detector. Using $e^+$ from radiactive sources the acceptance for at least
one of two 511~keV annihilation photons was determined to be
79(4)\% with 4.5(3)ns (FWHM) time resolution and 350(20) keV (FWHM)
energy resolution.
The transport path has 80(2)\% transmission and conserves
transverse spatial information of the decay vertex. It
can be reconstructed radially to 8.0(4)~mm and axially to 8.6(5)~mm (FWHM),
if in addition track parameters are used from the energetic $e^-$ detected in the
wire chambers.
The primary limitations on the resolution arise at high energies from
proportional chamber wire spacing and at
low energies from multiple scattering in the 1~mm carbon fiber beam tube.

During data taking, every 5 hours the M production yield was determined
at low beam rates (2$\cdot$10$^4$$\mu^+$/s)
using a method which is based on a model established
in independent dedicated experiments \cite{wood88}. It assumes M
production inside of the target where the $\mu^+$ stop. Their escape into vacuum
is governed by a random walk.
The number of atoms in the fiducial volume was determined mainly from
the distribution of time intervals between a beam counter signal
from the incoming muon and the detection of the atomic electron on the MCP (Fig. 3a).
The exponential background to these spectra was investigated using an Al foil target
and also negative muons stopped in SiO$_2$ powder (Fig. 3b).
In both cases there is no M production but secondary electrons
are released in a similar fashion as when a $\mu^+$ decays in the target
without M formation.
The procedure is in agreement with results obtained from a maximum likelihood
fit of the time distributions of M decay positions in four separate
spatial regions above the  production target.
Due to finite acceptances and efficiencies of the detector components
 $5.0(2)\cdot 10^{-3}$ of the incoming $\mu^+$ on average
were observed to decay as M atoms.
The M atom fraction in vacuum per incoming $\mu^+$ yields
2.0(1)\% which is a factor of 2.5 below the results of earlier
dedicated experiments where higher fractions were
achieved due to narrower beam momentum bites\cite{wood88}.
The SiO$_2$ targets were replaced twice a week, since the M
yield  deteriorated on a time scale of a week associated with the release
of H$_2$O molecules from the powder.

The final search result was obtained in three data taking periods
with a total duration of  6 month  distributed over 4 years (Table \ref{summary})
during which the sensitivity of the instrument was constantly improved.
In total $5.6(1)$$\cdot$10$^{10}$ M atoms
were investigated for $\overline{\rm M}$ decays.
Two major sources of potential background were identified:
(i) accidental coincidences of energetic $e^-$ produced by
Bhabha scattering of $e^+$ from M decays and scattered $e^+$
on the MCP and (ii) the allowed muon decay mode
$\mu \rightarrow e^+e^+e^-\nu_e \overline{\nu}_{\mu}$
with branching ratio 3.4$\cdot$10$^{-5}$
which could release an energetic $e^-$ and a low energy $e^+$
which are dtected and the
second $e^+$ escapes unobserved.
As the expected $e^+$ flight time was determined in M decays to
76(1)~ns with a 3.3(1)~ns wide (FWHM) distribution, a narrow coincidence window of
$\pm$4.5~ns was applied to suppress both processes.
Momenta of energetic $e^-$
were accepted between 15 and 90~MeV/c, which cuts 50(3)\% of
the Bhabha scattering events but only 8(2)\% of the Michel spectrum contents.
For the rare muon decay the bend of the $e^+$ transport system
spoils the vertex reconstruction because of the significantly higher
$e^+$ momenta. Positrons were required to deposit between
0.2 and 1.0 MeV in the CsI calorimeter and to be detected
within $\pm$6~ns of a hit on the MCP. The decay vertex had to be reconstructed
within $\pm$12~mm radially and was required to be not more than 3 cm upstream
of the SiO$_2$ target.

There was one event which passed all these required  criteria
within 3 standard deviations of each of
the relevant distributions which were
established in the regular M control measurements
prior to processing the search data
(Fig. \ref{res_mmb}).
The expected background due to accidental coincidences is 1.7(2) events
and was evaluated using data with a TOF 4.5~ns above the expected value.

The finite fiducial volume of diameter 9~cm and  length 10~cm
and the finite observation time are taken into account by
a correction factor of 0.82(2), which was determined in a Monte Carlo
simulation of the conversion process.
As a result an upper limit on the conversion
probability
of ${\rm P_{M\overline{M}} \leq 8.2\cdot 10^{-11}}/ {\rm S}_{\rm B}$
(90\% C.L.) was found.
The factor ${\rm S}_{\rm B}$ describes the suppression of the conversion
in external magnetic fields due to the removal of degeneracy between
corresponding levels in M and ${\rm \overline{M}}$.
It depends on the interaction type (Table \ref{magfield}).
The reduction is strongest for (V$\pm$A)$\times$(V$\pm$A) \cite{hori96,wong95}.
For these cases the traditionally quoted upper limit
for the coupling constant is
\begin{eqnarray*}
 {\rm G_{M\overline{M}}} &=& \rm G_F \cdot
 \sqrt{\frac{\rm P_{M\overline{M}} (0T)}{2.56\cdot10^{-5}}}\\
 &\leq& 3.0\cdot 10^{-3} {\rm G}_{\rm F} (90\%C.L.)
 \nonumber
\end{eqnarray*}
where ${\rm G}_{\rm F}$
is the weak interaction Fermi coupling constant.

This new result which exceeds previous bounds \cite{abel96} by a factor of 35
has some impact on speculative models.
A proposed ${\rm Z}_8$ model is ruled out with more than 4 generations of particles
where masses could be generated radiatively with heavy lepton seeding
\cite{wong94}.

A new lower limit of $m_{X^{\pm \pm}}$ $ \geq $ 2.6~TeV/c$^2$ $*g_{3l}$ (95\% C.L.)
on the masses of flavour diagonal bileptonic gauge bosons is found
in GUT models, which is well
beyond the value extracted from direct searches, measurements of the muon
magnetic anomaly or high energy Bhabha scattering
\cite{fuji94,cuyp96}, with $g_{3l}$ of order unity and depending on the details of the
underlying symmetry. For 331 models this translates into
$m_{X^{\pm \pm}}$ $ \geq $ 850~GeV/c$^2$, which excludes their minimal version
in which an upper bound of 800~GeV/c$^2$ has been extracted from electroweak
parameters \cite{fram97,fram97a}. The 331 model is only viable in an extended form
with, e.g. a Higgs octet
\cite{fram98}.

In the framework of R-parity violating supersymmetry \cite{moha92,halp93}
the bound on the coupling parameters could be lowered by a factor of 15
to $\mid \lambda_{132}\lambda_{231}^*\mid \leq 3 * 10^{-4}$ for assumed
superpartner masses of order 100 GeV/c$^2$ \cite{Moh98}.
Further the achieved level of sensitivity allows to narrow slightly the
interval of allowed
heavy muon neutrino masses
in minimal left-right symmetry \cite{herc92} (where a lower bound on
${\rm G_{M\overline{M}}}$ exists, if muon neutrinos are heavier than 35 keV/c$^2$)
to $\approx$ 40~keV/c$^2$ up to the present
experimental bound of 170~keV/c$^2$ \cite{assa94}.

In minimal left right symmetric
models, in which ${\rm {M\overline{M}}}$ conversion is allowed,
the process is intimately connected to
the lepton family number violating muon decay
$\mu^+ \rightarrow e^+ + \nu_{\mu} + \overline{\nu}_e$.
Within these models the limit achieved in this experiment
this decay is not an option
for explaining the excess neutrino counts in the LSND neutrino
experiment at Los Alamos  \cite{herc97}.

The consequences for atomic physics of M are such that the expected level
splitting in the ground state due to ${{\rm M} - \overline{{\rm M}}}$
interaction is below $1.5~{\rm Hz} / \sqrt{S_B}$ reassuring  the validity of
recent determinations of fundamental constants from atomic spectroscopy of the atom.

Further increased sensitivity to  ${{\rm M} \overline{{\rm M}}}$ conversion
could be expected, if the M production efficiency could be enhanced, or
from muon sources with significantly higher fluxes, e.g. at the front end
of a future muon collider.

This work is supported in part by
the German BMBF,
the Swiss Nationalfond,
the Russian FFR and a NATO research grant.
We thank P. Frampton, A. Halprin, P. Herczeg, R. Mohapatra and H.S. Pruys for
discussions on the implications of the measurements.
The excellent support we have received from PSI staff members was
essential for successful extended running.

\begin{table}[bht]
 \protect{
 \caption[]{
 \protect{ \label{summary}}Summary of three data taking series. The number of produced
 muonium atoms and the upper limits on the conversion probability in 0.1~T magnetic
 field are given. The probability in zero field can be obtained through division
 by the interaction dependent reduction factor $S_B$ of table \protect{\ref{magfield}}.
 }
         }
\begin{tabular}{crll}
 measurement & duration of  & number of
  & ${\rm P}_{{\rm M} \overline{{\rm M}}}$(0.1~T) \\
           & data taking  & muonium atoms & (90\%C.L.) \\
            \hline\\
  1  & 210 h  & $ 1.4(1) \cdot 10^{9}$  & $ 2.8 \cdot 10^{-9} $ \\
  2  & 230 h  & $ 2.3(2) \cdot 10^{9}$  & $ 1.8 \cdot 10^{-9} $ \\
  3  & 1290 h & $ 5.2(1) \cdot 10^{10}$ & $ 8.9 \cdot 10^{-11}$ \\
  all& 1600 h & $ 5.6(1) \cdot 10^{10}$ & $ 8.2 \cdot 10^{-11}$ \\
  \end{tabular}
\end{table}

%
%
\begin{table}[bht]
 \caption[]{
   Magnetic field correction factor  \protect{${\rm S}_{\rm B}$}
   of muonium-antimuonium
   conversion probability for muonium atoms with statistically
   populated ground states
   \protect{\cite{wong95,hori96}}.
          }
  \protect{\label{magfield}}
 \begin{tabular}{cccr}
 interaction type                                         & 2.6~$\mu$T&0.1~T&100~T\\
   \hline\\
   S S                                                     &0.75& 0.50 &0.50 \\
  P P                                                     &1.0 & 0.9  &0.50\\
  (V$\pm$A)$\times$(V$\pm$A) or (S$\pm$P)$\times$(S$\pm$P)&0.75& 0.35 &0.0\\
  (V$\pm$A)$\times$(V$\mp$A) or (S$\pm$P)$\times$(S$\mp$P)&0.95& 0.78 &0.67\\
  \end{tabular}
\end{table}

\begin{figure}[htb]
     \centering
      \epsfig{file=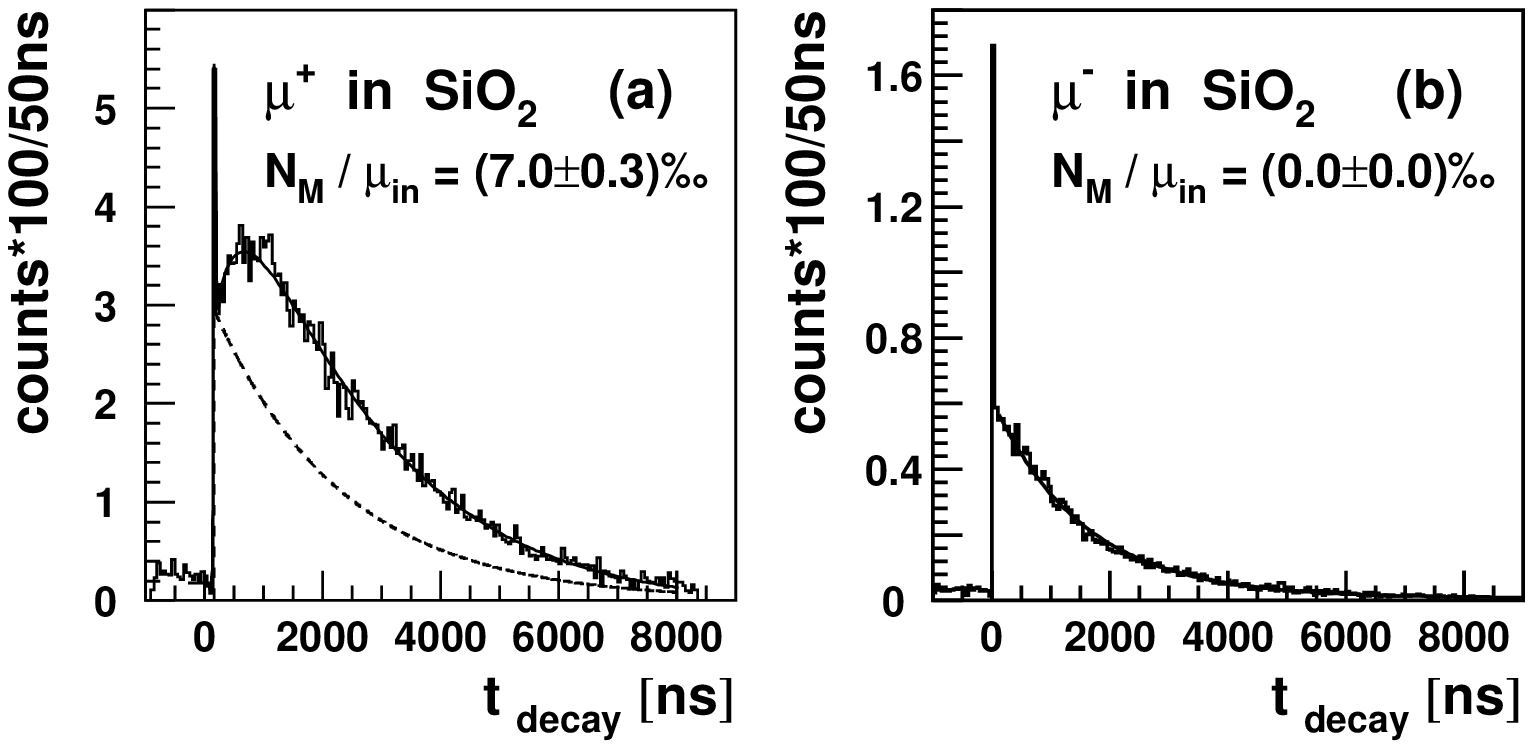,width=3.4in}
       \caption[]
       {\protect{\label{mproduction}}
       Muonium production was continually monitored using the
       characteristic time of flight distribution of atomic electrons (a).
       The indicated exponential background was verified
       by demonstrating that there is
       only an exponentially decaying background  for negative muons
       on SiO$_2$ powder (b).
       }
\end{figure}

\begin{figure}[htb]
       \begin{minipage}{1.5in}
        \centering{
        \hspace*{-0.8cm}
        \mbox{
        \epsfig{figure=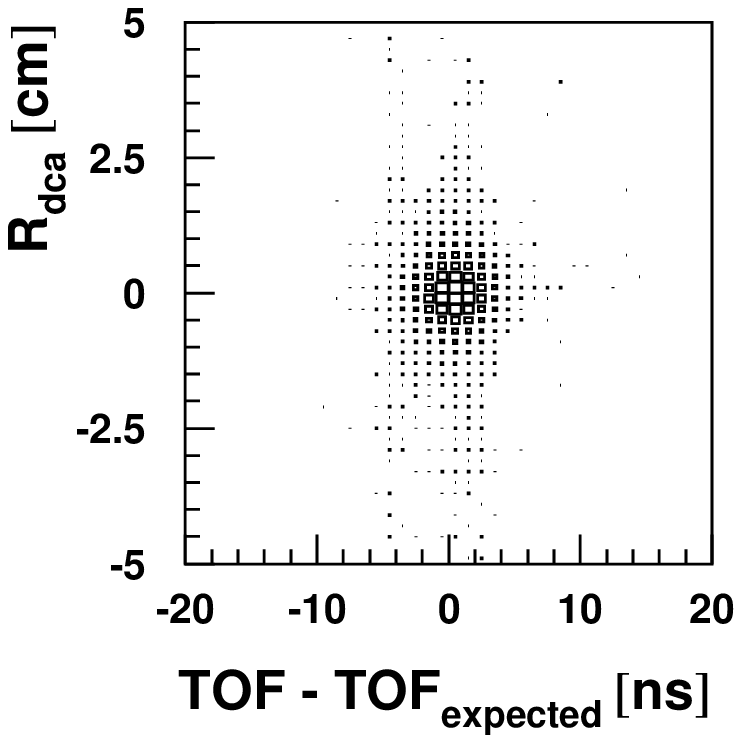,width=1.65in}
               }
              }
       \end{minipage}
       \hspace*{0.1cm}
       \begin{minipage}{1.5in}
        \centering{
        \hspace*{-0.8cm}
        \mbox{
        \epsfig{figure=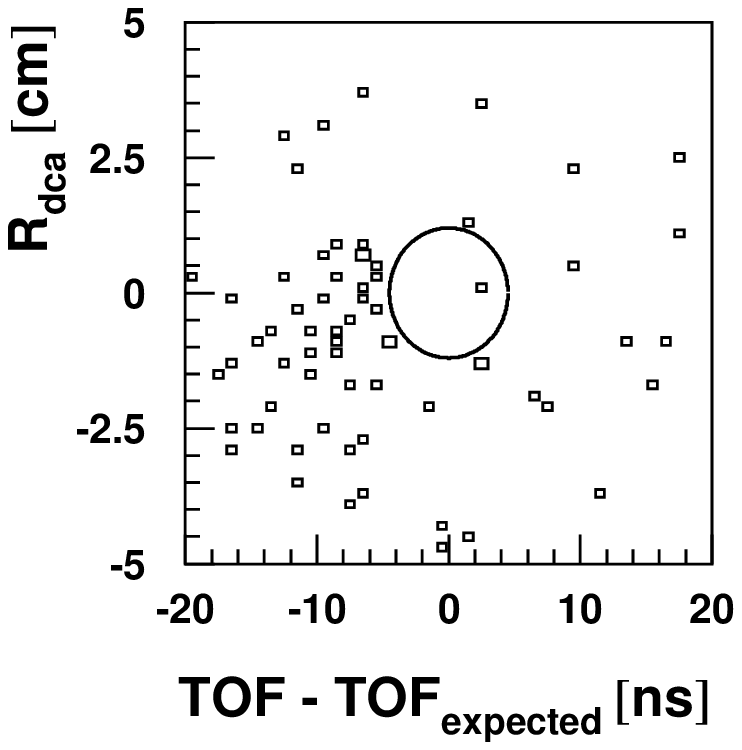,width=1.65in}
               }
             }
       \end{minipage}
       \centering\caption[]{
       \protect{\label{res_mmb}}
       The distribution of the distance of closest approach (R$_{dca}$)
       between a track from an energetic
       particle in the magnetic spectrometer and the back projection of the
       position on the MCP detector versus the time of flight (TOF)
       of the atomic shell particle, for a muonium measurement (left)
       and for all data recorded in the third data taking period of 1290 h
       while searching for antimuonium (right).
       One single event falls within a 3 standard deviations region
       of the expected TOF and R$_{dca}$ indicated by the ellipse.
       The events concentrated at low TOF
       and low R$_{dca}$correspond to the allowed decay
       $\mu \rightarrow 3e+2\nu$.
       }
       \end{figure}

\end{document}